\documentclass[10pt,onecolumn,oneside,notitlepage,final]{article}

\usepackage{slashbox}

\def \in #1 #2 {\int \limits_{#1}^{#2}}

\usepackage{amssymb}
\usepackage{amsmath}
\usepackage{graphicx} 
\usepackage{color}

\usepackage{amsmath}

\usepackage{pstricks}

\def\localinput#1{{
  \renewcommand{\documentclass}[2][dummy]{}
  \renewcommand{\usepackage}[2][dummy]{}
  \renewenvironment{document}{}{}
  \def\jobname{#1}
  \input{#1}
}}

\def\sla#1{\ooalign{\hfil\hspace{-0.1ex}\raise.2ex\hbox{$\not \phantom{#1}$}\hfil\crcr  $#1$}}

\begin{document}

\begin{titlepage}

\date{}

\begin{flushright}
RUP-10-2
\end{flushright}
\vspace{12ex}
\begin{center}
 \huge{Determination of the Higgs CP property in Hadron Colliders }

\vspace{3ex}

\large{ Akihiro Matsuzaki\footnote{akihiro@rikkyo.ac.jp}

\small \textsl{ Center For Educational Assistance, Shibaura Institute Of Technology,} 
 \\ \small \textit{307 Fukasaku, Minuma-ku, Saitama-shi, Saitama 337-8570 JAPAN  }
 }
\\[4mm]

\large{ Hidekazu Tanaka\footnote{tanakah@rikkyo.ac.jp}\\
\small \textsl{ Department of Physics, Rikkyo University,}
 \\ \small \textit{Nishi-ikebukuro, Toshima-ku Tokyo, Japan, 171  }
 }

\vspace{3ex}


\end{center}

\begin{abstract}
    We propose three ways to determine the CP eigenvalue of the Higgs boson at the hadron collider as follows: 
    1. We determine the Higgs CP eigenvalue from the production cross section which is affected by the CP eigenvalue of the Higgs boson.
    2. We adopt the CP selection rules to determine the Higgs CP eigenvalue.
    3. We determine the CP property by the momentum distribution of the decay products of the Higgs boson.
    Our methods can be applied for a wide range of the Higgs mass.

\end{abstract}

\end{titlepage}

\section{Introduction}

    The Higgs boson search is being started at the Large Hadron Collider (LHC).
    The energy will be larger and the events are collected hereafter.
    It is the most important mission of the LHC to discover the Higgs bosons which is the last Standard Model (SM) particle to be discovered.
    However, the higgs sector may not be the SM one.
    For examples, the Two Higgs Doublet Model (2HDM) has five kinds of Higgs bosons and the Supersymmetric (SUSY) SM also has.
    To understand this, we have to study the Higgs property in detail. 
    The mass of Higgs bosons is determined when the bosons are discovered and so is the electromagnetic (EM) charge.
    How about the CP property?
    The CP eigenvalue of the SM Higgs boson is even.
    On the other hand, the SUSY models, the composite Higgs models, and the techni color model have the CP-odd Higgs bosons \cite{higgs hunters guide} and \cite{9411426}.
    To determine the parity of the Higgs bosons, a method is already proposed in Ref. \cite{9404280}.  
    If the Higgs mass is larger than 182 (160) GeV, it can decay into $ZZ$ ($W^+W^-$). 
    Then, some $Z$ ($W$) decay into the lepton pair. 
    Also, if the Higgs mass is heavier than 350 GeV, Higgs bosons can decay into $t\bar t$.
    Then, some $t$ decay into charged lepton by the semileptonic decay. 
    The momentum distribution of these leptons tells us the parity of the Higgs boson.
    If the Higgs mass is lighter, the Higgs cannot decay into $Z Z$, $W^+ W^-$, or $t\bar t$, and Higgs bosons mainly decay into $b \bar b$.

    The main subject of this paper is to show how to determine the CP property of Higgs, model independently.
    We give three ways to identify the CP-odd Higgs bosons as follows.
    
    First, the CP odd Higgs interaction differs from the CP even one.
(    The Higgs CP property makes difference between their interactions.)
    However, if the Higgs mass is lighter, the Higgs decay modes are restricted as explained before.
    Then, we have an interest in the creation process.
    
    Second, the selection rules restrict several decay modes.
    Concretely, we study the Higgs decays into $b$ hadrons since the dominant part of the Higgs bosons which mass is lighter than $160$ GeV decay into $b \bar b$ quark pair.

    Third, for heavier Higgs bosons, we consider the Higgs decay modes, $\phi\to ZZ$, $\phi\to W^+ W^-$, $\phi\to t \bar t$.
    These modes tell us the parity of the final states.
    The C parity of these final states are even as we explain later.
    Then, we can determine the CP eigenvalue of the Higgs boson.

    This paper is organized as follows.
    In Section \ref{section 2}, we show how to identify the CP-odd Higgs bosons using the creation process. 
    In Section \ref{section 3}, we apply this to the 2HDM.
    In Section \ref{section 4}, we apply this to the Minimal Supersymmetric SM (MSSM).
    In Section \ref{section 5}, we try another way to determine the CP property of Higgs bosons, i.e. using the selection rules.
    In Section \ref{section 6}, we consider CP property of the heavier Higgs bosons, which decay into $ZZ$, $W^+ W^-$, or $t \bar t$. 
    In Section \ref{section 7}, we present our conclusion on the determination of Higgs CP property.

\section{Higgs creation process}\label{section 2}

    We define the neutral Higgs bosons $\phi=\{S,A\}$, where $S$ and $A$ are the CP even and odd Higgs bosons, respectively.
    In the 2HDM and MSSM, for example, $S=\{h,H\}$ are the light and heavy CP even Higgs bosons, respectively.
    
    We consider the CP conserving Lagrangian here since if the CP symmetry of the Lagrangian are violated, the CP eigenvalue of the Higgs bosons is not a good quantum number.
    
    The Higgs boson-massive gauge boson-massive gauge boson interaction ($\phi$-$V$-$V$) is explained by the effective Lagrangian written in the Lorentz and gauge invariant operators at the leading order as
\begin{align} \begin{split}
\mathcal{L}&\ni 
  D^\mu S D_\mu \langle S \rangle 
+\frac{1}{4}\frac{g_{\mathrm{odd}}}{M}A \epsilon_{\mu \nu \rho \sigma} F^{\mu \nu}F^{\rho \sigma}   
\\
&\ni 
  g_{\mathrm{even}}M S V^\mu V_\mu
 +\frac{g_{\mathrm{odd}}}{M} A    \epsilon_{\mu \nu \rho \sigma} \partial^\mu V^\nu \partial^\rho V^\sigma,   
\end{split} \end{align}
    where $D_\mu$ is the covariant derivative; $F^{\mu \nu}$ is the field strength tensor;  $\epsilon_{\mu \nu \rho \sigma}$ is the totally antisymmetric tensor; $M$ is a constant which has the mass dimension; $\langle S \rangle$ means the vacuum expectation value of $S$; and $g_{\mathrm{even}}$ and $g_{\mathrm{odd}}$ are the coupling constants of $S$ and $A$, respectively.
    We note here the $\epsilon_{\mu \nu \rho \sigma}$ parity is odd.

    The second term cannot be renormalized and we need a loop diagram.    
    Since the loop diagram has two gauge couplings, comparing to $S$-$V$-$V$ diagram, $A$-$V$-$V$ coupling will be suppressed more than the factor $\alpha_W^2$ (for MSSM, see Ref. \cite{p2907-1}), 
    where $\alpha_W\simeq1/30$ is the weak fine structure constant. 
    On the other hand, the $\phi$-$t$-$t $ diagrams are naturally the same order of magnitude for $\phi=S$ and $\phi=A$.

    Then, especially in the early stage of the experiment or with low luminosity, we can say that if the Higgs boson decays (does not decay) into $VV$, it is CP even (odd) Higgs.
    However, if the Higgs boson mass is lower than about $160$ GeV, this does not work since we consider $Z$ or $W$ bosons as $V$.
    Therefore, we should study the creation rather than the decay.
    $A$ cannot be created by the vector boson fusion while $S$ can.
    If we can reconstruct the $\phi$ invariant mass from $t \bar t \to \phi$ process while (and) we cannot (can) reconstruct from $VV\to \phi$, then we can say that $\phi=A$ $(S)$, respectively. 

    Actually, in 2HDM and MSSM, it works.
    However, if $g_{\mathrm{even}}$ is small for any reasons, we cannot determine the CP eigenvalue of $\phi$ since both $S$ and $A$ cannot produced via the vector boson fusion.

\section{$\phi$-$V$-$V$ and $\phi$-$f$-$f$ Couplings in the 2HDM}\label{section 3}

    In this section, we consider the Type I and II 2HDM for instance.
    Table \ref{Table 1} suggests that the $\phi$-$f$-$f$ and $\phi$-$V$-$V$ type coupling constants normalized by the SM prediction in type I and II 2HDM, where $f$ is a fermion.
    Generally, the interaction of $A$ is different from that of $h$ and $H$.
    Also, the interaction of type I Higgs bosons are different from that of type II Higgs bosons.

\begin{table}[h]
 \caption{The coupling constants normalized by the SM prediction. }
 \begin{center}\label{Table 1}
  \begin{tabular}{|c||c|c|c||c|c|c|}
    \hline
    &\multicolumn{3}{|c||}{Type I}&\multicolumn{3}{|c|}{Type II}  \\
\cline{2-7}
            &  $h$   &  $H$   &  $A$   &  $h$   &  $H$   &  $A$   \\
    \hline
    \hline
  & & & & & &  \\[-10pt]
\shortstack{ $\phi$-$t$-$t$,\\[-1pt] $\phi$-$c$-$c$} &\shortstack{ $\cos\alpha $\\[-1pt]$\overline{\sin\beta}$ }  &  \shortstack{$\sin\alpha $\\[-1pt]$ \overline{\sin\beta}$}    & \shortstack{1\\[-1pt]$\overline {\tan\beta}$}   &  \shortstack{$\cos\alpha$\\[-1pt]$\overline{\sin\beta}$}    &  \shortstack{$\sin\alpha$\\[-1pt]$\overline{\sin\beta}$}    & \shortstack{1 \\ $\overline{\tan\beta}$}    \\[0pt]
    \hline
  & & & & & &  \\[-10pt]
\shortstack{ $\phi$-$ b$-$b$,\\[-1pt] $\phi$-$ \tau$-$\tau$} &\shortstack{ $\cos\alpha $\\[-1pt]$\overline{\sin\beta}$ }  &  \shortstack{$\sin\alpha $\\[-1pt]$ \overline{\sin\beta}$}    & \shortstack{1\\[-1pt]$\overline {\tan\beta}$}   &  \shortstack{$\sin\alpha$\\[-1pt]$\overline{\cos\beta}$}    &  \shortstack{$\cos\alpha$\\[-1pt]$\overline{\cos\beta}$}    & \shortstack{\phantom{.} \\ \raisebox{1.3ex}{$\tan\beta$} }   \\[0pt]
    \hline
  & & & & & &  \\[-10pt]
\shortstack{ $\phi$-$ Z$-$Z$,\\ $\phi $-$W$-$W$}           & \raisebox{1.3ex}{$\sin(\beta-\alpha)$ }  &\raisebox{1.3ex}{ $\cos(\beta-\alpha)$}   & \raisebox{1.3ex}{0}   &\raisebox{1.3ex}{ $\sin(\beta-\alpha)$}   &\raisebox{1.3ex}{ $\cos(\beta-\alpha)$}   &\raisebox{1.3ex}{ 0}   \\[1pt]
    \hline
  \end{tabular}
 \end{center}
\end{table}

\begin{table}[h]
 \caption{Same as Table \ref{Table 1} for $\beta=\alpha$.}
 \begin{center}\label{Table 2}
  \begin{tabular}{|c||c|c|c||c|c|c|}
    \hline
    &\multicolumn{3}{|c||}{Type I}&\multicolumn{3}{|c|}{Type II}  \\
\cline{2-7}
            &  $h$   &  $H$   &  $A$   &  $h$   &  $H$   &  $A$   \\
    \hline
    \hline
  & & & & & &  \\[-10pt]
\shortstack{ $\phi$-$t$-$t$,\\[-1pt] $\phi$-$c$-$c$} &\shortstack{ $1 $\\[-1pt]$\overline{\tan\beta}$ }  &  \raisebox{1.3ex}{1}    & \shortstack{1\\[-1pt]$\overline {\tan\beta}$}   &  \shortstack{1\\[-1pt]$\overline {\tan\beta}$}     &  \raisebox{1.3ex}{1}     & \shortstack{1 \\ $\overline{\tan\beta}$}    \\[0pt]
    \hline
  & & & & & &  \\[-10pt]
\shortstack{ $\phi$-$ b$-$b$,\\[-1pt] $\phi$-$ \tau$-$\tau$} &\shortstack{ $1$\\[-1pt]$\overline{\tan\beta}$ }  &  \raisebox{1.3ex}{1}     & \shortstack{1\\[-1pt]$\overline {\tan\beta}$}   &  \raisebox{1.3ex}{$\tan\beta$}     &  \raisebox{1.3ex}{1}     & \raisebox{1.3ex}{$\tan\beta$}    \\[0pt]
    \hline
  & & & & & &  \\[-10pt]
\shortstack{ $\phi$-$ Z$-$Z$,\\ $\phi $-$W$-$W$}           & \raisebox{1.3ex}{$0$ }  &\raisebox{1.3ex}{$1$}   & \raisebox{1.3ex}{0}   &\raisebox{1.3ex}{$0$}   & \raisebox{1.3ex}{1}   &\raisebox{1.3ex}{0}   \\[1pt]
    \hline
  \end{tabular}
 \end{center}
\end{table}

\begin{table}[h]
 \caption{Same as Table \ref{Table 1} for $\beta=\alpha+\pi/2$.}
\begin{center}\label{Table 3}
  \begin{tabular}{|c||c|c|c||c|c|c|}
    \hline
    &\multicolumn{3}{|c||}{Type I}&\multicolumn{3}{|c|}{Type II}  \\
\cline{2-7}
            &  $h$   &  $H$   &  $A$   &  $h$   &  $H$   &  $A$   \\
    \hline
    \hline
  & & & & & &  \\[-10pt]
\shortstack{ $\phi$-$t$-$t$,\\[-1pt] $\phi$-$c$-$c$} &  \raisebox{1.3ex}{1}  &\shortstack{ $1 $\\[-1pt]$\overline{\tan\beta}$ }    & \shortstack{1\\[-1pt]$\overline {\tan\beta}$} &  \raisebox{1.3ex}{1}   &  \shortstack{1\\[-1pt]$\overline {\tan\beta}$}         & \shortstack{1 \\ $\overline{\tan\beta}$}    \\[0pt]
    \hline
  & & & & & &  \\[-10pt]
\shortstack{ $\phi$-$ b$-$b$,\\[-1pt] $\phi$-$ \tau$-$\tau$}  &  \raisebox{1.3ex}{1}  &\shortstack{ $1$\\[-1pt]$\overline{\tan\beta}$ }    & \shortstack{1\\[-1pt]$\overline {\tan\beta}$}  &  \raisebox{1.3ex}{1} &  \raisebox{1.3ex}{$\tan\beta$}          & \raisebox{1.3ex}{$\tan\beta$}    \\[0pt]
    \hline
  & & & & & &  \\[-10pt]
\shortstack{ $\phi$-$ Z$-$Z$,\\ $\phi $-$W$-$W$}         &\raisebox{1.3ex}{$1$}    & \raisebox{1.3ex}{$0$ }   & \raisebox{1.3ex}{0}   &\raisebox{1.3ex}{$1$}   & \raisebox{1.3ex}{0}   &\raisebox{1.3ex}{0}   \\[1pt]
    \hline
  \end{tabular}
 \end{center}
\end{table}

    If we know that the two mixing angles $\alpha$ and $\beta$ are not the special values where $\beta-\alpha=0$ or $\pi/2$ (see Table \ref{Table 2} and \ref{Table 3}), we apply the result in previous section.
    However, If we do not know $\alpha$ and $\beta$, we have to consider that possibility.
    We explain below what we can say when we discover one, two, and three Higgs candidates, respectively.

\subsection{One Higgs candidate }
   
    At first, the LHC may discover only one scalar particle.
    We point out the following:     
\begin{itemize}
  \item If the scalar particle is not created by the vector boson fusion, which corresponds to the $\phi$-$ V$-$V$ coupling constant normalized by the SM prediction $g_{\phi VV}=0$, it is not the SM-like particle, however we cannot determine which it is.
  \item If the coupling constant is $g_{\phi VV}\not=0$, it is not $A$.
\end{itemize}

\subsection{Two Higgs candidates }
    Next, we consider that we have two resonances which are the Higgs candidates.
\begin{itemize}
  \item     If their coupling constants are both $g_{\phi VV}=0$, then one of them is $A$. Also, the undiscovered one is the SM-like Higgs.
  \item     If one has $g_{\phi VV}=0$ and the other has $g_{\phi VV}=1$, the former is undetermined but the latter is the SM-like Higgs.
  \item    If both of them have $g_{\phi VV}\not=0$, they are $h$ and $H$.
  \item    If one has $g_{\phi VV}=0$ and the other has $g_{\phi VV}\not=0,1$, the former is $A$ and the latter is undetermined.

\end{itemize}

\subsection{Three Higgs candidates}
    The last, we consider that we have three resonances which are the Higgs candidates.
\begin{itemize}

\item When one of $g_{\phi VV}$ is $1$ and the others are $0$, we can make the following statements:

    \begin{itemize}
    \item If the couplings are $g_{\phi VV}=\{1,0,0\}$ in order of increasing mass, then $\phi$ are $\{h,H,A\}$ or $\{h,A,H\}$.
    \item If the couplings are $g_{\phi VV}=\{0,1,0\}$ in order of increasing mass, then $\phi$ are $\{h,H,A\}$ or $\{A,h,H\}$.
    \item If the couplings are $g_{\phi VV}=\{0,0,1\}$ in order of increasing mass, then $\phi$ are $\{h,A,H\}$ or $\{A,h,H\}$.
    \end{itemize}

\item When one of the $g_{\phi VV}$ is $0$ and others are not $0$ or $1$, the former one is $A$, the latter two are $h$ and $H$.

\item If the two mixing angles $\beta-\alpha=0$, $H$ becomes SM-like Higgs boson and the interaction of $h$ is the same as that of $A$ (see Table \ref{Table 2}).
    On the other hand, if $\beta-\alpha=\pi/2$, $h$ becomes SM-like Higgs boson and the interaction of $H$ is the same as that of $A$ (see Table \ref{Table 3}). 
   
\end{itemize}

\subsection{$\phi$-$f$-$f$ couplings in the 2HDM}

    In the 2HDM, the Higgs bosons couple to the fermion pair.
    For instance, Higgs to $t\bar t$ pair coupling has a relation
\begin{align} \begin{split}
g_{htt}^2+g_{Htt}^2-g_{Att}^2
=
\left(\frac{\cos\alpha}{\sin\beta}\right)^2+\left(\frac{\sin\alpha}{\sin\beta}\right)^2-\left(\frac{\cos\beta}{\sin\beta}\right)^2
=1,
\end{split} \end{align}
    where $g_{\phi tt}$ are the $\phi$-$t$-$t$ coupling constant normalized by the SM top Yukawa coupling constant. 
    This relation suggests that if we find the two Higgs candidates and the sum of squares of their normalized coupling constants is smaller than 1, one of these candidates must be $A$.
    We can say the same thing for $\phi$-$b$-$b$ and $\phi$-$\tau$-$\tau$ couplings.
    
    If $\alpha-\beta=0,$ or $\pi/2$, we cannot get any information more than the information from $\phi$-$V$-$V$ coupling.

\section{MSSM Prediction}\label{section 4}

    We here consider the MSSM prediction in the following two situations.

    First, we apply large $\tan\beta$. 
\begin{table}[h]
 \caption{The MSSM coupling constants normalized by the SM prediction, for $\tan\beta\gg 1$. }
 \begin{center}\label{Table 4}
  \begin{tabular}{|c||c|c|c|}
    \hline
              &  $h$   &  $H$   &  $A$   \\
    \hline
    \hline
 $\phi$-$ t$-$t$, $\phi$-$ c$-$c$         & $\cos\alpha$    & $\sin\alpha$ & 0   \\[1pt]
    \hline
 $\phi $-$b$-$b$, $\phi $-$\tau$-$\tau$   & $\sin\alpha \tan\beta$    & $\cos\alpha \tan\beta$ & $\tan\beta$    \\[1pt]
    \hline
 $\phi$-$ Z$-$Z$, $\phi$-$ W$-$W$          & $\cos\alpha$   & $\sin\alpha$   & 0   \\[1pt]
    \hline
  \end{tabular}
 \end{center}
\end{table}
    In this situation,
    $g_{htt}^2\simeq g_{hVV}^2\simeq \cos^2\alpha$, $g_{Htt}^2\simeq g_{HVV}^2\simeq \sin^2\alpha$, and $g_{Att}^2\simeq g_{AVV}^2 \simeq 0$ as written in Table \ref{Table 4}.
    The CP odd Higgs cannot be created from the vector boson fusion and also from $\phi$-$t$-$t$ vertex.
    However, it is created from $\phi$-$b$-$b$ vertex.

    Next, if $m_h=98$ GeV, $m_H=120$ GeV, and $h$-$Z$-$Z$ interaction is too weak to discover in the LEP as in Ref. \cite{hep-ph/0609076}, where $m_h$ and $m_H$ are the masses of $h$ and $H$, respectively, then $g_{hZZ}^2\simeq 0.1$ and $g_{HZZ}^2\simeq 0.9$.
    Thus, if a Higgs candidate is found and 
    the $\phi$-$V$-$V$ coupling is smaller than 10 percent of the SM prediction, 
    we can say that it must be a CP odd Higgs.

\section{Selection Rules}\label{section 5}

    As you saw above, the first method does not work in some situations.
    Then we try another way.
    We here study what we can say about the Higgs CP property using the selection rules.
    Some decay modes are forbidden by the selection rules, and then we can determine the initial state CP eigenvalue. 
    For example, $\eta$ meson cannot decays into two pions, which state is CP even.
    This fact means that $\eta$ is CP odd particle.
    We apply this to the Higgs decaying into two $b$ hadrons.

    However, apparently, we have some problems as follows
    (actually, these problems can be solved and we will see them later): 
\begin{enumerate}
\item When the Higgs boson decays into quarks, they mostly do not produce two-body final state but jets. Then the CP selection rules may not work well.
\item Also, we have to consider the Higgs decay modes, not only into $B \bar B$ but also $B \bar B^*$, $B^* \bar B$, $B^* \bar B^*$, $B \bar B^{**}$, $\Lambda_b \bar \Lambda_b$, etc. 
These modes give a similar signal since these hadrons are in the jets. 
Then, the selection rules may be hidden.
\end{enumerate}

\subsection{Jets}
    The jets created from the Higgs decays are divided in two types.

    One is the jet which is created from the $b$-hadrons.
    These are mainly made of $\pi^{\pm,0}$.
    As explained later, for example, the $S\to B \bar B$ decays are allowed, on the other hand, the $A\to B \bar B$ decays are forbidden by CP conservation.
    When we denote these interactions by the effective Lagrangian which is constructed by Higgs and hadrons, the $A$-$B$-$B$ vertices themselves are forbidden by CP conservation.  
    Therefore, we do not have to care of the sequential interactions which create many pions.

    The other is a gluon jet which energy is larger than $B$ meson mass.
    This jet is interpreted to be created before the $b$ quark constructs hadrons.
    However, these jets are suppressed by the fine structure constant of strong interaction which is perturbative in this energy region. 
    Moreover, we can eliminate the three jet events except for the case where the gluon jet merges into one of the quark jets.
    Then, this kind of jets is not serious either and we can divide the Higgs bosons by their CP properties.

\subsection{ $B^{(*,**)} \bar B^{(*,**)}$ Decays}

    The lighter Higgs mainly decays into $b \bar b$.
    They finally become hadrons.

    In the two meson final states, they consist of $B,B^*,B^{**}$ and their antiparticles, where 
    $B^{**}$ are $B_0^*,B_1',B_1,B_2^*$, which have $J^{P}=0^+,1^+,1^+,2^+$, respectively \cite{fermilab-thesis-2008-84}.
    Also, $B$ and $B^*$ have $J^P=0^-,1^-$, respectively.

    The $44.3\%$ of $\bar b$ quark fragments into $B^*$ and the $31.9\%$ into $B^{**}$ \cite{cer-002272784.ps}.
    The $B$ production rate is one third of $B^*$ production rate since $B^*$ is spin 1 and has three degrees of freedom \cite{9710505}, \cite{ZPhysC74-437}. 
    The production rate ratio of $B_0^*$, $B_1'$, $B_1$, $B_2^*$ is $1:3:3:5$ corresponding to their spin degrees of freedom.
    Then, we suppose that the production rate ratio of $B$, $B^*$, $B_0^*$, $B_1'$, $B_1$, $B_2^*$ for one degree of freedom is $r_0:r_0:r_1:r_1:r_1:r_1$, where $r_0:r_1=44.3/3:31.9/12=1772:319$.

    When the spin-0 particle decays into the particle-antiparticle pair, the total spin $s$ and the angular momentum $L$ of final state are the same.
     Then, the C-parity $C$ is 
\begin{align} \begin{split}
C=(-1)^{s+L}=(-1)^{2L}=+1.
\end{split} \end{align}
    This relation does not care if the final-state particles are the fermions or bosons.

    We particularly see about the final states, $B\bar B$, $B \bar B^*$, $B^* \bar B$, and $B^* \bar B^*$ as examples. 

\subsubsection{$B\bar B$}

    The C parity of $B\bar B$ final state is $+1$ as already explained.

    The intrinsic parity of $B$ and $\bar B$ are both $-1$.
    $\phi$, $B$, and $\bar B$ are all spin-0 particles.
    Then, the angular momentum in the final state $L=0$ and the parity 
\begin{align} \begin{split}
P=(-1)(-1)(-1)^L=+1.
\end{split} \end{align}
    Then, the CP eigenvalue $CP=+1$.
    We conclude that $A\to B \bar B$ decay mode is forbidden by the CP selection rules.

\subsubsection{$B \bar B^*$, $B^* \bar B$}

    In $B \bar B^*$ and $B^* \bar B$ states, we cannot define the C parity since $\bar B^*$ is not the antiparticle of $B$.
    However, almost 100\% of $\bar B^*$ decays into $\bar B$ with a 46 MeV photon \cite{ZPhysC74-413}.
    Then, the successive decay process is 
\begin{align} \begin{split}
\phi \to B+ \bar B^* \to B+ \bar B +\gamma,
\end{split} \end{align}
    where we can define C parity in its final state $B\bar B\gamma$.

    First, we consider the parity.
    The final state has a relation
\begin{align} \begin{split}
|L-s|\le J\le L+s,
\end{split} \end{align}
    where $L$, $s$, and $J$ are the orbital angular momentum, total spin, and total angular momentum, respectively.
    In $\phi \to B \bar B^*$ process, $J=0$ since $\phi$ is a spin-0 particle, and $s=1$, then $L=s=1$.
    Therefore, the parity in $B \bar B^*$ state becomes $P=(-1)(-1)(-1)^L=-1$.
    In $\bar B^* \to \bar B +\gamma$ process, parity cannot violated since the related interactions are only EM and even strong ones.
    Then, the parity in the final state $ B \bar B \gamma$ is $P=-1$.

    Generally, there are possible $2L+1$ states labeled by the $z$-component of the total angular momentum $L_z$ for each $L$.
    However, $L_z \not=0$ states are forbidden in two body decay and only one state $L_z=0$ is possible for each $L$.

    Next, we consider the C parity.
    In this process, the Higgs boson decays into $b \bar b$ quark pair.
    Subsequently, it becomes $B\bar B\gamma$.
    This can be written as 
\begin{align} \begin{split}
\langle  B \bar B \gamma  |\mathcal{S}|\phi\rangle
=\langle  B \bar B \gamma  |  \mathcal{S}_{\mathrm{strong,\ EM} }    | b\bar b \rangle \langle b\bar b|\mathcal{S}_{\mathrm{Yukawa}} |\phi\rangle,
\end{split} \end{align}
    where  $\mathcal{S}$, $\mathcal{S}_{\mathrm{strong,\ EM}}$, and $\mathcal{S}_{\mathrm{Yukawa}}$ are the S-Matrix of the theory, strong and EM interaction, and Yukawa interaction, respectively.
    Here, the C parity in $b\bar b$ state is $C=+1$ and $ \mathcal{S}_{\mathrm{strong,\ EM}}  $ cannot violate the C parity, thus the C parity in  $B \bar B \gamma$ state becomes $C=+1$. 

    Therefore, the CP eigenvalue in the $ B \bar B \gamma$ state becomes $CP=-1$.
    Then, $S\to B\bar B^*$ and $S\to B^* \bar B$ decays and $S$-$B^*$-$B$ effective vertex are forbidden by CP selection rules.

\subsubsection{$B^* \bar B^*$}

    The C parity of $B^*\bar B^*$ final state is $+1$ as explained before.

    We consider the parity of the final state.
    Now, $s=L$ since $J=0$, and two spin-1 particles give the total spin $s=2,1,0$, then  $L=2,1,0$, respectively.
    If $L=2,0$, the parity $P=(-1)(-1)(-1)^L=+1$ and $CP=+1$.
    On the other hand, if $L=1$, the parity $P=(-1)(-1)(-1)^L=-1$ and $CP=-1$.
    Then, there are two CP even states and one CP odd state.

\subsection{Other modes}

    As summarized in Table \ref{Table 5}, there are many mesonic two-body decay modes. 
    We calculate the CP eigenvalues for these modes.
    In these modes, $B_0^*$ and the half of $B_2^*$ decay into $B$,
    Also, $B_1'$, $B_1$, and the half of $B_2^*$ decay into $B^*$ \cite{fermilab-thesis-2008-84}, \cite{0705.3229}, \cite{0711.0319}.
    $B^*$ decays into $ B + \gamma$.
    Then, each mode contains 0, 1, or 2 photons.

    In the two-body baryonic decay modes, they consist of $\Lambda_b$, $\Sigma_b$, $\Sigma_b^*$, $\Xi_b$, $\Xi_b^*$, $\Omega_b$ and their antiparticles,
    where they have $I(J^P)=0(\frac{1}{2}^+)$, $1(\frac{1}{2}^+)$, $1(\frac{3}{2}^+)$, $\frac{1}{2}(\frac{1}{2}^+)$, $\frac{1}{2}(\frac{3}{2}^+)$, $0(\frac{1}{2}^+)$, respectively.
    Considering the EM charge, isospin, and spin conservation, only  
    $\Lambda_b \bar \Lambda_b$, $\Sigma_b^{(*)} \bar \Sigma_b^{(*)}$, $\Xi_b^{(*)} \bar \Xi_b^{(*)}$, and $\Omega_b \bar \Omega_b$ modes are allowed.
    Thus, the C parity is even, 
    and each final state has one parity even state and one odd state.
    Then, the number of states in the baryonic decay modes is the same for CP even and odd Higgs.
    Anyway, they are kinematically forbidden to produce $B^*$ or $\bar B^*$ mesons and then the photon from them.

\begin{table}[htbp]
 \caption{The mesonic two-body decay modes. 
    The second column is the number of $\gamma$ from $B^*$ or $\bar B^*$ in decayed particles.
    The third and fourth columns are the number of CP even and odd states of each mesonic mode.
    The fifth and sixth columns are the production probability in mesonic modes for CP even and odd Higgs, respectively.
}
 \begin{center}\label{Table 5}
  \begin{tabular}{|c|c|c|c|c|c|}
    \hline
  & & \multicolumn{2}{|c|}{CP}&\multicolumn{2}{|c|}{probability}  \\
\cline{3-6}
 mode  & number  &   &   & CP even & CP odd \\
                 &of $\gamma$ & even & odd  &  $(\times (3r_0^2 + 8r_1^2 \hfill$  &  $(\times (r_0^2 + 4r_1^2+  c_c r_0^2 $\\
                 &            &      &      &  $+ 8c_c r_0 r_1 + 5 c_c r_1^2) )$ &  $ + 6 c_c r_0 r_1 +7 c_c r_1^2))$\\
    \hline
    \hline
 $B B$           & 0   &  1  & 0   & $r_0^2 $        & $0$ \\
    \hline
 $B B^*$         & 1   &  0  & 1   & $0$             & $ c_c r_0^2$  \\
    \hline
 $B B_0^*$       & 0   & 1   &  0  & $c_c r_0 r_1$   &  $0$  \\
    \hline
 $B B_1'$        &  1  & 1   & 0   & $c_c r_0 r_1$  & $0$ \\
    \hline
 $B B_1 $        &  1  &  1  &  0  & $c_c r_0 r_1$  &  $0$  \\
    \hline
 $B B_2^* $      & 1,0 & 0   & 1   & $0$             & $c_c r_0 r_1$  \\
    \hline
    \hline
 $B^* B^* $      &  2  &  2  &  1  & $ 2r_0^2  $     & $ r_0^2  $  \\
    \hline
 $B^* B_0^* $    & 1   &  1  & 0   & $c_c r_0 r_1$  &0  \\
    \hline
 $B^* B_1' $     & 2   &  1  & 2   & $c_c r_0  r_1$ & $2c_c r_0  r_1  $ \\
    \hline
 $B^* B_1 $      &  2  &  1  &  2  & $c_c r_0  r_1$ & $2c_c r_0  r_1  $  \\
    \hline
 $B^* B_2^* $    & 1,2 & 2  &  1  & $2c_c r_0 r_1$ & $c_c r_0  r_1  $  \\
    \hline
    \hline
 $B_0^* B_0^*$   & 0   & 1   &  0  & $r_1^2     $    &0  \\
    \hline
 $B_0^* B_1'$    & 1   &  0  &  1  & $0$             & $c_c r_1^2$ \\
    \hline
 $B_0^* B_1$     &  1  &  0  &  1  & $0$             & $c_c r_1^2$  \\
    \hline
 $B_0^* B_2^*$   & 0,1 & 1   &  0  & $c_c r_1^2$    & $0$ \\
    \hline
    \hline
 $B_1' B_1' $    &  2  &  2  &  1  &  $2r_1^2 $      &  $r_1^2 $  \\
    \hline
 $B_1' B_1 $     &  2  & 2   &  1  &$2c_c r_1^2 $    &$c_c r_1^2 $  \\
    \hline
 $B_1' B_2^* $   & 1,2 & 1   & 2  & $c_c r_1^2$    & $2c_c r_1^2 $ \\
    \hline
    \hline
 $B_1 B_1$       & 2   & 2   & 1   & $2r_1^2 $       & $r_1^2 $  \\
    \hline
 $B_1 B_2^*$     & 1,2 & 1   & 2  & $c_c r_1^2$    & $2c_c r_1^2 $ \\
    \hline
    \hline
 $B_2^* B_2^*$   &0,1,1,2& 3& 2  & $3r_1^2 $      & $2r_1^2 $  \\
    \hline
  \end{tabular}
 \end{center}
\end{table}

    The $90.9\%$ of $b\bar b$ pairs fragment into mesonic modes,
    Also, the $9.1\%$ of $b\bar b$ pairs fragment into baryonic modes.
    Then, the fragmentation probability of the modes in which CP is odd and the number of $\gamma$ from $ B^* $ or $\bar B^* $ decay is 0 is  
\begin{align} \begin{split}\label{event rate}
\frac{\frac{1}{2}c_c r_0 r_1 +\frac{2}{4} r_1^2}
{r_0^2 + 4r_1^2 +  c_c r_0^2 + 6 c_c r_0 r_1 +7 c_c r_1^2 
} \times 0.909+ 0.091=0.122,    
\end{split} \end{align}
    where the factor $c_c=2$ comes from the charge conjugate mode.  
    We give Table \ref{Table 6} by similar calculation.

\begin{table}[htbp]
 \caption{The fragmentation probability of the final states which have 0, 1, and 2 $\gamma$'s, respectively from $ B^* $ or $\bar B^* $ decay for each CP eigenvalue of the Higgs boson.}
 \begin{center}\label{Table 6}
  \begin{tabular}{|c|c|c|}
    \hline
   number of $\gamma$    & CP even   & CP odd   \\
    \hline
    0   &  29.5 \% & 12.2 \%  \\  
    \hline
    1   &  22.3 \% & 42.0 \%  \\  
    \hline
    2   &  48.2 \% & 45.8 \%  \\  
    \hline
  \end{tabular}
 \end{center}
\end{table}
    The ratio of the zero photon event number over the one photon event number for CP even Higgs and CP odd Higgs are
\begin{align} \begin{split}
\frac{29.5}{22.3}&= 1.32  \hspace{2.5em} \mathrm{(for\ CP\ even)},\\ 
\frac{12.2}{42.0}&= 0.291 \hspace{2em}   \mathrm{(for\ CP\ odd )}, 
\end{split} \end{align}
    respectively.
    Then, we can determine the CP eigenvalue of the Higgs boson if we detect the $\gamma$ from $B^*$ and $\bar B^*$.

    If an energetic baryon such as $p$, $\bar p$, $n$, $\bar n$, $\Lambda$, or $\bar \Lambda$ in each $b$-jet is identified, we can guess that this event is a baryonic mode. 
    If we eliminate them from the analysis, the event rates derived in Eq. (\ref{event rate}) is modified and shown in Table \ref{Table 7}.
\begin{table}[htbp]
 \caption{As Table 6, but the baryonic modes can be eliminated.}
 \begin{center}\label{Table 7}
  \begin{tabular}{|c|c|c|}
    \hline
   number of $\gamma$    & CP even   & CP odd   \\
    \hline
    0   &  22.4 \% & 3.42 \%  \\  
    \hline
    1   &  24.5 \% & 46.2 \%  \\  
    \hline
    2   &  53.0 \% & 50.4 \%  \\  
    \hline
  \end{tabular}
 \end{center}
\end{table}
    Then, the ratio of the zero photon event number over the one photon event number for CP even Higgs and CP odd Higgs are
\begin{align} \begin{split}
\frac{22.4}{24.5}&= 0.914  \hspace{2.5em} \mathrm{(for\ CP\ even)},\\ 
\frac{3.42}{46.2}&= 0.0740 \hspace{2em}   \mathrm{(for\ CP\ odd )}, 
\end{split} \end{align}
    respectively.

\section{$\phi \to$ $Z Z$, $W^+ W^-$, $t \bar t$}\label{section 6}

    We here show how to determine the Higgs CP eigenvalue if the Higgs mass is heavy.

\subsection{$\phi \to$ $Z Z$, $W^+ W^-$}
    If the Higgs mass is larger than 160 GeV, the $\phi \to W^+ W^-$ mode is kinematically allowed,
    also if it is larger than 182 GeV, $\phi \to Z Z$ mode is kinematically allowed.
    As explained in section \ref{section 2}, the $A$-$V$-$V$ coupling is highly suppressed, and the CP odd Higgs does not decay into the vector boson pair.
    Then, we can say that if the Higgs boson decays (does not decay) into $VV$, it is CP even (odd) Higgs.

\subsection{$\phi \to$ $t \bar t$}

    For instance, when $\tan\beta$ is large in the 2HDM and MSSM, the Higgs boson mainly decays into $b\bar b$ even if the Higgs mass is larger than 350 GeV \cite{9803257}.
    On the other hand, when $\tan\beta$ is not so large and the Higgs mass is larger than 350 GeV, it mainly decays into $t \bar t$ pair.
    
    In  $\phi\to t \bar t$, the top quark decays weakly into lighter particles before hadronization.
    Here, the C parity of $t \bar t$ pair state is even.
    Therefore, if we can determine the parity in this state, we can also determine the CP eigenvalue of $\phi$. 

\begin{figure}[ht]
  \begin{center}
    \includegraphics[keepaspectratio=true,height=70mm]{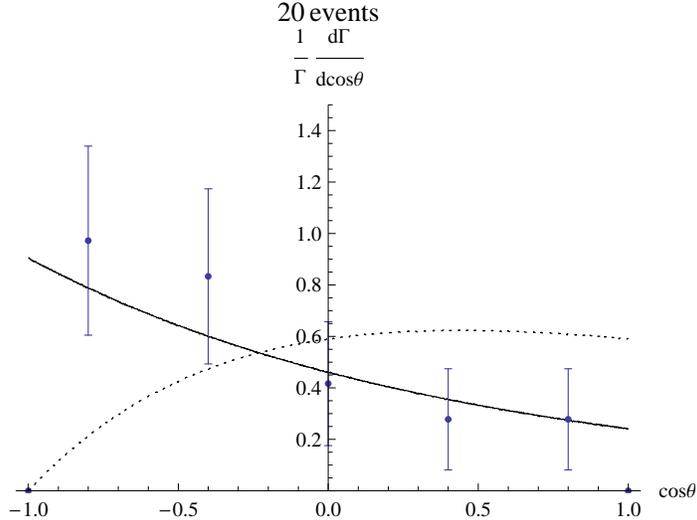}
  \end{center}
  \caption{
    The opening angle distribution in $\phi$ rest frame for $m_\phi=400$ GeV.
    The solid line is for $D_2/D_1=1$, which corresponds to parity (and also CP) even. 
    The dotted line is for $D_2/D_1=-1$, which corresponds to parity (CP) odd.
    The data points with error bars show the result of Monte Carlo simulation for $D_2/D_1=1$ in a sample of 20 events.
    This simulation results $\frac{D_2}{D_1}=1.4\pm0.7$.
}
  \label{fig:H-t-bart-1-mH400GeV-20events.eps}
\end{figure}

\begin{figure}[ht]
  \begin{center}
    \includegraphics[keepaspectratio=true,height=70mm]{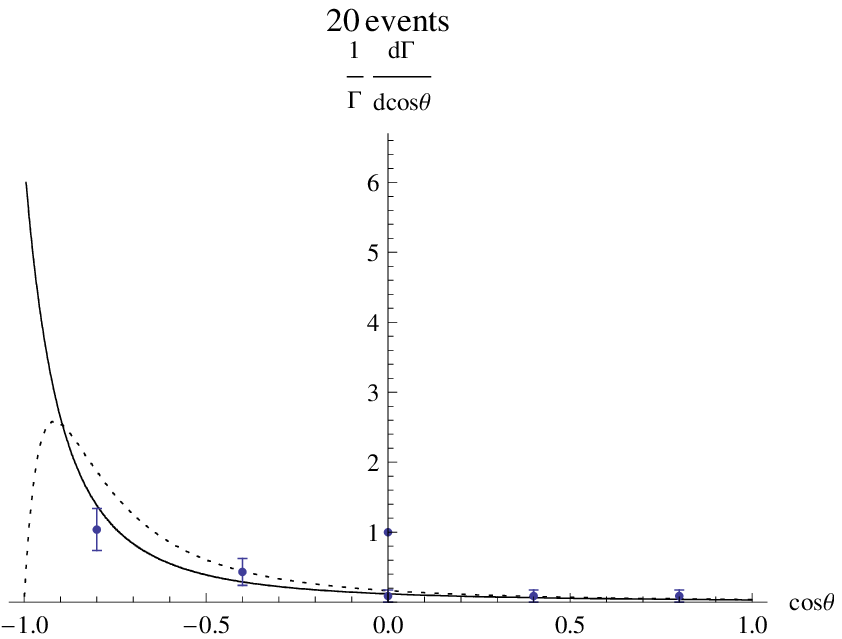}
  \end{center}
  \caption{
    The opening angle distribution for $m_\phi=1000$ GeV.
    Others are the same as in Fig. \ref{fig:H-t-bart-1-mH400GeV-20events.eps}.
    This simulation results $\frac{D_2}{D_1}=1.8\pm 1.1$.
    }
  \label{fig:H-t-bart-1-mH1000GeV-20events.eps}
\end{figure}

    To determine the parity of $t \bar t$, we detect the $\ell^+ \ell^-$ opening angle distribution of $\phi \to t \bar t \to b W^+ + \bar b W^-  \to b \ell^+ \nu_\ell + \bar b \ell^- \bar \nu_\ell$ decay process as Ref. \cite{9404280}. 
    The results of Monte Carlo simulation are depicted in Figs. \ref{fig:H-t-bart-1-mH400GeV-20events.eps} and \ref{fig:H-t-bart-1-mH1000GeV-20events.eps}, and
\begin{align} \begin{split}
\frac{D_2}{D_1}&=1.4\pm 0.7\ \ \  \mathrm{for}\ m_\phi=400\  \mathrm{GeV}, \ \phi=S,\ 20\ \mathrm{events},  \\   
\frac{D_2}{D_1}&=1.8\pm 1.1\ \ \  \mathrm{for}\ m_\phi=1000\ \mathrm{GeV}, \ \phi=S,\ 20\ \mathrm{events},   
\end{split} \end{align}
    where $m_\phi$ is the mass of $\phi$ and the parameter $D_2/D_1$ is defined in Ref. \cite{0904.4375}
    $D_2/D_1$ should be $1$ when $t \bar t$ parity (and CP) is even and $-1$ when $t \bar t$ parity (CP) is odd.

    The number of events we can detect depends on the collider luminosity and the cross section.
    For instance, in the MSSM \cite{9803257} with $\tan \beta=1.5$ and $m_A=400$ GeV, about 10 events of  
\begin{align} \begin{split}
pp\to   &A + t \bar t\\
&\hspace{0.3em}{}^\lfloor \hspace{-0.5em} \longrightarrow  t \bar t   \to b \ell^+ \nu_\ell + \bar b \ell'{}^- \bar \nu_{\ell'}
\end{split} \end{align}
    process are generated for one year at the LHC, which luminosity will be $10^{34} \mathrm{cm^{-2}s^{-1}}$.
    Here, $\ell$ and $\ell'$ are the leptons.    
    Ref. \cite{9507411} says that 
    the signal events are reduced to 50 \% and the QCD background is reduced to about 10 \% of the signal events by the cuts.

\section{Conclusion}\label{section 7}

    We proposed three ways to determine the CP eigenvalue of the Higgs candidate in the hadron collider.
    First, we showed that the CP even and odd Higgs bosons are created in different process.
    The CP even Higgs can be created in the vector boson fusion.
    On the other hand, the CP odd Higgs cannot be produced in that process.
    Then, we can determine the CP eigenvalue of the Higgs candidate from the creation process.
    This can be used for light Higgs boson which mass is lighter than about $160$ GeV.
    However, this method cannot work if $S$-$V$-$V$ coupling happens to be small.   
    Actually, as we seen in Section \ref{section 3}, in 2HDM or MSSM, we cannot determine the CP eigenvalue of Higgs candidate if $\beta=\alpha$, $\alpha+\pi/2$.

    Second, we applied the CP selection rules to $\phi\to b\bar b$ decay modes.
    This method can also apply to the light Higgs candidates.
    We suggested that the CP even Higgs candidate tends to decay with no photon from $B^*$ or $\bar B^*$ rather than the CP odd Higgs candidate.
    This is useful to determine the CP eigenvalue of the Higgs candidates.
    Especially, if we can divide the mesonic modes from the baryonic modes, 
    the ratio of the number of photon is definitely different between CP even and odd Higgs candidates.

    Last, we showed that the opening angle distribution, which determines the parity of the Higgs candidate, also determines the CP eigenvalue of them.
    We showed the Monte Carlo simulation for $m_S=400$ GeV and $1000$ GeV.
    For each situation, we can determine the CP eigenvalue of the Higgs candidate in a sample of 20 events.

    These three methods allow us to determine the CP eigenvalue of the Higgs candidates with a wide range of mass spectrum.

    If the data do not suggest either CP even or odd in the analysis of selection rules and $\phi\to t\bar t $ decay, it means CP violation  
    since we do not suppose any special new physics in their analysis.

    The LHC Higgs events are coming soon.
    We hope that these methods will work well to determine the CP eigenvalue of the Higgs candidates.

\section{Acknowledgements}
    This work is in part supported by Department of Physics and Research Center for Measurement in Advanced Science in Rikkyo University. 


\end{document}